\documentclass[12pt,a4paper]{article}

\usepackage{amssymb}
\usepackage{setspace}
\usepackage{indentfirst}

\usepackage{graphicx}

\begin{document}

\begin{titlepage}
\begin{center}

\vspace*{25mm}

\begin{spacing}{1.7}
{\LARGE\bf
 Self-interacting dark matter
 and core formation in field low-surface-brightness galaxies}
\end{spacing}

\vspace*{25mm}

{\large
Noriaki Kitazawa
}
\vspace{10mm}

Department of Physics, Tokyo Metropolitan University,\\
Hachioji, Tokyo 192-0397, Japan\\
e-mail: noriaki.kitazawa@tmu.ac.jp

\vspace*{25mm}

\begin{abstract}
Dark matter may play an important role in galaxy formation
 through its non--trivial properties.
For example,
 self--interacting dark matter
 may contribute to the formation of the widely observed core structures in galaxies.
However,
 galaxy formation is a complex process,
 and such core structures can also arise from baryonic effects
 within the cold dark matter framework.
To clarify the role of dark matter self--interactions,
 it is necessary to study systems
 that evolve without significant baryonic disturbances.
Low--surface--brightness galaxies in the field,
 which are gravitationally isolated and have evolved with minimal external influence,
 are suitable candidates for this purpose.
Since these galaxies typically contain only a small amount of baryonic matter,
 strong baryonic effects are not expected in their evolutionary history.
In this study,
 we assume that these galaxies decoupled from proto--clusters at high redshift.
Based on this assumption,
 we set initial conditions and estimate the time required for core formation,
 which we compare with the time corresponding to the redshift of proto--clusters.
 We examine five low--surface--brightness galaxies in the field
 and three observed proto--clusters at redshifts $z=2.45$, $7.66$ and $7.88$.
Our analyses, based on order--of--magnitude estimates without numerical simulations,
 excludes a self--interaction cross section of $\sigma/m = 1 \, {\rm cm^2/g}$,
 while $\sigma/m = 0.1 \, {\rm cm^2/g}$ is favored.
This result is consistent with constraints
 derived from the shapes of present--day cluster cores.
\end{abstract}

\end{center}
\end{titlepage}

\doublespacing

\section{Introduction}
\label{sec:introduction}

The non--trivial nature of dark matter as an elementary particle
 may be required to understand its role
 within the standard model of particle physics.
Several problems
 in small--scale cosmological structure formation with cold dark matter
 suggest such a non--trivial nature.
In particular core--cusp problem in dwarf galaxies
 \cite{Flores:1994gz,Moore:1994yx} is of interest,
 because baryonic effects appear insufficient
 to transform the typical cold dark matter density profile, namely
 the Navarro--Frenk--White (NFW) profile \cite{Navarro:1996gj},
 into a cored profile.
If dark matter self--interactions have an appropriate strength,
 core formation may be explained
 through gravothermal collapse \cite{Lynden-Bell:1968eqn,Lynden-Bell:1980xip}
 without spoiling the success of cold dark matter on large scales.

The dynamics of self--interactions
 in the evolution from an NFW profile to a cored profile
 have been studied in detail
 using the gravothermal fluid approximation \cite{Pollack:2014rja}
 and N--body simulation \cite{Choquette:2018lvq}.
Simulations for more realistic galaxies, although still neglecting baryonic effects,
 have also been performed \cite{Koda:2011yb,Rocha:2012jg,Elbert:2014bma}
 demonstrating the possibility of core formation.
However, it has been pointed out \cite{Rocha:2012jg} that
 major mergers of galaxies can suppress core formation,
 since it is the slow thermal process.
Furthermore,
 it has been argued that baryonic effects alone may account for core formation
  in cold dark matter halos (see \cite{Tulin:2017ara} and references therein).
Therefore,
 to isolate the effects of dark matter self--interactions,
 it is necessary to study galaxies
 that are free from strong baryonic processes and major mergers.
Low--surface--brightness galaxies (LSBGs) in the field provide such an environment.

LSBGs contains very small amount of baryonic matter compared to dark matter.
Most of the baryonic component is in the form of neutral hydrogen,
 and these systems are thought to have experienced little star formation,
 with few supernova explosions or strong stellar winds.
Field LSBGs evolve in relative isolation,
 largely unaffected by interactions with other galaxies.
This makes them suitable systems
 for studying slow thermal processes such as gravothermal collapse.
Observationally,
 many field LSBGs exhibit cored dark matter distributions,
 inferred from rotation curves measured using the 21cm or Lyman--$\alpha$ lines.
In \cite{DiPaolo:2019eib},
 rotation curves of $72$ LSBGs were analyzed in detail
 and fitted with a phenomenological model.
In the present study
 we select five field LSBGs from this sample 
 that show smooth rotation curves
 without prominent features such as kinks or overshoots.

We assume that
 LSBGs were originally part of proto--clusters and latter migrated into the field.
This is a strong assumption and
 should be tested by future simulations of structure formation.
We consider three observed proto--clusters
 at redshifts $z=2.45$ (about $10.8 \simeq 11$ Gyr ago) \cite{Diener:2014efa},
 $z=7.66$ (about $12.80 \simeq 13$ Gyr ago) \cite{Laporte:2022} and
 and $z=7.88$ (about $12.83 \simeq 13$ Gyr ago) \cite{Morishita:2022}
\footnote{
The age of universe is $13.797 \simeq 14$ Gyr \cite{ParticleDataGroup:2024cfk}.
}.
These studies identify
 the proto--clusters as regions
 with a significant concentration of galaxies at a given redshift.
They provide estimates of
 the number of member galaxies, their typical masses
 and the size of over--density of the proto--clusters.
Using the rotation curves fits from \cite{DiPaolo:2019eib} 
 together with results from Millennium simulation
 \cite{Springel:2005nw,Kitzbichler:2006ec,Henriques:2011xn},
 we determine the initial conditions of dark matter halos
 in terms of NFW profile parameters.
We then apply the results in \cite{Pollack:2014rja}
 to assess whether the dark matter cores can form by the present time
 in the selected five field LSBGs.

In the next section
 we briefly review gravothermal collapse
 within the gravothermal fluid approximation following \cite{Pollack:2014rja}. 
In section \ref{sec:LSBG}
 we review observation of the proto--cluster at $z=2.45$ \cite{Diener:2014efa}
 and describe our method for setting the initial NFW profiles of the LSBGs.
Our analysis is based on order--of--magnitude estimates,
 and more precise results will require future numerical simulations.
We consider the initial conditions corresponding to three observed proto--clusters.
In section \ref{sec:core}
 we discuss core formation through gravothermal collapse in the five field LSBGs.
We find that the self-interaction cross section
 should be $\sigma/m \sim 0.1 \, {\rm cm^2/g}$,
 consistent with constraints from the shapes of present--day cluster halos.
The final section summarizes our conclusions.

\section{Gravothermal collapse by self-interacting dark matter}
\label{sec:SIDM}

Consider a sphere of non--interacting particles
 in which gravity and pressure (arising from kinetic motion) are balance.
If the system is in virial equilibrium,
 the virial theorem gives a relation between
 the total kinetic energy $T$ and potential energy $V$ as $T=-V/2$.
Since the total energy $E$ is given by $E=T+V$,
 we obtain $E=-T$, implying a negative specific heat.
If the particles move toward center
 due to an input kinetic energy (heat),
 they gain a larger amount of negative potential energy,
 and the total energy decreases.
Now consider the case in which particles have self--interactions:
 one particle may be scattered inward while another moves outward. 
The inner particle gains
 kinetic energy and much more negative potential energy
 to balance the stronger gravitational force,
 while the outer particle loses kinetic energy
 in response to the weaker gravitational field.
As a result, In total energy decreases,
 and the repetition of this process leads to a concentration of
 particles toward the center.
This provides an intuitive picture of gravothermal collapse.

We employ the gravothermal fluid approximation
 \cite{Lynden-Bell:1980xip,Balberg:2002ue,Koda:2011yb},
 which provides five equations describing gravothermal collapse
 in a spherically symmetric system:
\begin{equation}
 \frac{\partial M}{\partial r} = 4 \pi r^2 \rho,
\label{mass}
\end{equation}
\begin{equation}
 \frac{\partial \left( \rho \nu^2 \right)}{\partial r}
  = - \frac{GM\rho}{r^2}, 
\label{hydrostatic}
\end{equation}
\begin{equation}
 \frac{L}{4\pi r^2} = - \kappa \frac{\partial T}{\partial r},
\label{heat-propagation}
\end{equation}
\begin{equation}
 \frac{\partial L}{\partial r}
  = - 4 \pi r^2 \rho \nu^2
    \left(\frac{\partial}{\partial t}\right)_M \ln \frac{\nu^3}{\rho}.
\label{heat-gradiant}
\end{equation}
The first equation determines the relation
 between the enclosed mass $M(r,t)$ and density $\rho(r,t)$.
The second equation represents hydrostatic equilibrium,
 involving the velocity dispersion $\nu^2(r,t)$ and the gravitational constant $G$.
The third equation describes heat transport
 where $L(r,t)$ is the inward heat flux,
 $\kappa$ is the thermal conductivity
 (which depends on the self--interaction cross section),
 and the temperature $T$ defined as $T=m\nu^2/k_B$,
 with $m$ the dark matter mass and $k_B$ the Boltzmann constant.
The last equation describes the radial variation of the heat flux;
 the time derivative is taken at constant enclosed mass
 ({\it i.e.}, a Lagrangian derivative).
The dependent variables are $M$, $\rho$, $\nu$, and $L$, all functions of $r$ and $t$.

There are two important length scales and one characteristic timescale in this system.
The first is the gravitational scale height (or Jeans length),
\begin{equation}
 H \equiv \sqrt{\frac{\nu^2}{4 \pi G \rho}}
\end{equation}
 which characterizes the size of the system.
The second is the mean free path,
\begin{equation}
 \lambda \equiv \frac{1}{\rho\sigma},
\end{equation}
 where $\sigma$ denotes the self--interaction cross section par unit mass.
The third is the relaxation time,
\begin{equation}
 t_r \equiv \frac{1}{a\rho\nu\sigma}
\end{equation}
 with $a=\sqrt{16/\pi}$.
This is the typical timescale over which
 the system returns to equilibrium after a perturbation.
The thermal conductivity $\kappa$
 can be estimated using these three scales as
\begin{equation}
 \kappa \simeq
  \frac{3}{2} k_B \frac{\rho}{m}
  \cdot \frac{1}{t_r} \left[ \frac{1}{H^2} + \frac{1}{\lambda^2} \right]^{-1}.
\end{equation}
As a result, the heat transport equation becomes
\begin{equation}
 \frac{L}{4\pi r^2}
  \simeq - \frac{3}{2} \nu \sigma
           \left[ \sigma^2 + \frac{4\pi G}{\rho\nu^2}\right]^{-1}
           \frac{\partial \nu^2}{\partial r}.
\end{equation}
More precise expressions, including numerical factors of order unity,
 are given in \cite{Pollack:2014rja,Balberg:2002ue}.

To solve these equations, we must specify initial conditions at a given time.
Assuming that the self--interaction strength
 is not so large enough to significantly alter halo formation in the early universe,
 we adopt the NFW profile as the initial density distributions:
\begin{equation}
 \rho_i(r) = \frac{\rho_s}{(r/r_s) \cdot (1+r/r_s)^2},
\label{NFW}
\end{equation}
 where $\rho_s$ and $r_s$ are constants.
The initial value of $M$, $\nu$, and $L$,
 can be obtained by integrating
 eqs.(\ref{mass}),\, (\ref{hydrostatic}) and (\ref{heat-propagation}).
An important quantity is the initial relaxation time,
\begin{equation}
 t_{r,0} \equiv \frac{1}{a\rho_s\nu_s\sigma},
\label{initial-tr}
\end{equation}
 where $\nu_s \equiv \sqrt{4 \pi G \rho_s r_s^2}$.

A detailed analysis has been done in \cite{Pollack:2014rja} whose results we follow.
They show that
 the time required to form a cored profile from an initial NFW profile at
 $t = (10 \sim 50) \times  t_{r,0}$ (see Figure 2 in \cite{Pollack:2014rja}).
After this, the core structure remains nearly unchanged up to $250 \times t_{r,0}$,
 while at $\sim 450 \times t_{r,0}$ the core undergoes collapse into a black hole
 (see Figure 3 in \cite{Pollack:2014rja}).
We define the reference timescale for core formation as
\begin{equation}
 t_{\rm core} \equiv 30 \times t_{r,0}.
\label{time-core-form}
\end{equation}
Since gravothermal collapse is a slow thermal process,
 the core profile remains nearly unchanged up to $8 \times t_{\rm core}$,
 and black hole formation occurs only after a much longer time,
 around $\sim 15 \times t_{\rm core}$.

\section{Field low-surface brightness galaxies from proto-clusters}
\label{sec:LSBG}

As described in section \ref{sec:introduction}
 we assume that LSBGs migrated into the field
 after initially belonging to a proto--cluster.
In this section
 we briefly explain about observations of proto--cluster
 and explain how we determine the initial conditions of LSBGs.

We first consider
 the proto--cluster observed at $z=2.45$ by \cite{Diener:2014efa}.
In this study,
 a proto--cluster is identified
 through the detection of $11$ galaxies within a radius of $1.4$ Mpc (physical)
 with spectroscopically determined redshifts in the range $2.439 < z < 2.453$.
The overdensity of the proto--cluster is estimated as
\begin{equation}
 \delta = \frac{\rho_{\rm cl} - \rho_{\rm field}}{\rho_{\rm field}} \simeq 10,
\end{equation}
 where $\rho_{\rm cl}$ is the average density of the proto--cluster
 and $\rho_{\rm field} \equiv \rho_m(z=2.45)$ is the background matter density,
 with $\rho_m(z)=\rho_{\rm crit,0}\,\Omega_m(1+z)^3$
 is the background matter density at $z$
\footnote{
Here, $\rho_{\rm crit,0}=3H_0^2/8 \pi G$
 and we adopt $\Omega_m=0.3$ and $H_0=70$ km/s/Mpc, assuming a flat universe.
}.
By compare with the Millennium simulation,
 it has been shown that such proto--clusters evolve into Virgo- or Coma--like clusters.
In this context
 the typical mass of the member galaxies is estimated as $M \simeq 10^{12} M_\odot$,
 corresponding to the typical virial radius of about $100$ kpc.

We now describe how to determine the NFW parameters in eq.(\ref{NFW}).
If the radius and mass of a member galaxy,
 $R_{\rm mem}$ and $M_{\rm mem}$, are known, eq.(\ref{mass}) gives
\begin{equation}
 M_{\rm mem}
 = \int_0^{R_{\rm mem}} dr \, 4 \pi r^2 \, \rho(r,t)
 = 4 \pi \rho_s R_{\rm mem}^3
   \left[ \ln(1+R_{\rm mem}/r_s) - \frac{R_{\rm mem}/r_s}{1+R_{\rm mem}/r_s} \right].
\label{mass-explicit}
\end{equation}
Since the radius $R_{\rm mem}$ and $M_{\rm mem}$ are not directly observable,
 we assume that the size and total mass do not change significantly
 during the evolution of LSBGs in the field.
We therefore approximate them
 by the core radius $R_c$ and dynamical mass $M_{\rm dyn}$
 obtained from fits of present--day LSBGs in \cite{DiPaolo:2019eib}.
This provides a relation between $\rho_s$ and $r_s$.

Next, we relate the characteristic density $\rho_s$
 to the proto--cluster density $\rho_{\rm cl} = (1 + \delta) \rho_{\rm field}$.
This can be written as
\begin{equation}
 \rho_{\rm cl} = \frac{1}{V} \sum_{i=1}^N M_i,
\end{equation}
 where $V$ is the volume of the proto--cluster,
 $M_i$ is the mass of each number galaxy, and $N$ is the number of members.
The volume can be decomposed as
\begin{equation}
 V = V_{\rm void} + \sum_{i=1}^N V_i,
\end{equation}
 where $V_i$ is the volume of each galaxy
 and $V_{\rm vold}$ is the volume between galaxies. 
From this, we obtain
\begin{equation}
 \sum_{i=1}^N V_i \rho_i = V \rho_{\rm cl}
 = \sum_{i=1}^N V_i \left( 1 + \frac{V_{\rm void}}{NV_i} \right) \rho_{\rm cl},
\end{equation}
 with $\rho_i \equiv M_i/V_i$.
We then assume that the typical density $\rho_s$ is given by
\begin{equation}
 \rho_s = \left( 1 + \frac{V_{\rm void}}{NV_s} \right) \rho_{\rm cl}
 \equiv A \, \rho_{\rm cl},
\end{equation}
 where $V_s$ is a typical galaxy volume
 and $A$ is a characteristic constant of the proto--clustar.
For the proto--clustar considered here,
 taking $V$ as a sphere of radius $1.4$ Mpc,
 $V_s$ as a sphere of radius $100$ kpc, and $N=11$,
 we estimate
\begin{equation}
 \rho_s = A \, \rho_{\rm cl} \qquad \mbox{with} \qquad A \simeq 100.
\end{equation}
Solving eq.(\ref{mass-explicit}) then provides as estimate of $r_s$.
In this way, we determine the initial conditions of LSBGs at $z=2.45$.

Next we consider the proto--cluster observed at $z=7.66$
 using the gravitational lensing \cite{Laporte:2022}.
The proto--cluster volume is approximated as a sphere of radius $60$ kpc,
 with $N=8$ member galaxies the overdensity $\delta \simeq 4$.
The galaxy masses range 
 from $2 \times 10^{10} M_\odot$ to $6 \times 10^{11} M_\odot$,
 and the total mass is estimated as $3.34^{+0.59}_{-0.50} \times 10^{11} M_\odot$.
Taking a typical value of the mass of $5\times10^{10}M_\odot$
 and a corresponding virial radius of $14$ kpc from the Millennium simulation,
 we estimate $A \simeq 10$.

Finally, we consider the proto--cluster at $z=7.88$
 observed via gravitational lensing \cite{Morishita:2022}.
The proto--cluster volume is approximated as a sphere of radius $60$ kpc,
 with $N=15$ galaxies and overdensity $\delta \simeq 24$.
The mass of a representative bright galaxy is estimated as $7\times10^{10}M_\odot$,
 corresponding to a virial radius of $15$ kpc from the Millennium simulation.
In this case, we obtain $A \simeq 5$.

We emphasize that
 these estimates are based on order-of-magnitude arguments
 and involve several assumptions.
More precise analyses will require
 high--resolution cosmological simulations of proto--clusters at high redshifts.
Such simulations are particularly important
 for understanding the origin of field galaxies, especially LSBGs.
Proto--clusters may serve as a sources of field galaxies,
 since the their member galaxies are not yet in viral equilibrium
 and may be ejected into the field under certain conditions.

\section{Core formations}
\label{sec:core}

We consider five representative field LSBGs
 observed in \cite{deBlok:1996jib} and further analyzed in \cite{DiPaolo:2019eib},
 as summarized in Table \ref{five-LSBGs}.
All five galaxies exhibit smoothly rising rotation curves
 without prominent features such as kinks, overshoots,
 suggesting that their evolution
 has not been significantly affected by strong baryonic processes.
This is supported by the low values of the ratio $M_{\rm HI}/M_{\rm dyn}$,
 where $M_{\rm HI}$ denotes the total mass of neutral hydrogen,
 namely the baryonic component of these galaxies.
No central black holes have been observed in these galaxies,
 although they exhibit cored dark matter profiles.

\begin{table}[t]
\begin{center}
\large
\begin{tabular}{|l|c|c|c|l|} \hline
 name    & $M_{\rm dyn}$ & $M_{\rm HI}/M_{\rm dyn}$ & $R_c$ & shape \\ \hline\hline
 F561-1  & $10^{9.66}M_\odot$  & 0.177 & $25$kpc & bulge + faint disk \\ \hline
 F563-V1 & $10^{9.00}M_\odot$  & 0.283 & $14$kpc & faint bar \\ \hline
 F571-V1 & $10^{10.13}M_\odot$ & 0.049 & $21$kpc & faint ragged \\ \hline
 F574-1  & $10^{10.43}M_\odot$ & 0.072 & $34$kpc & faint spiral \\ \hline
 F574-2  & $10^{9.50}M_\odot$  & 0.293 & $33$kpc & core + faint disk \\ \hline
\end{tabular}
\caption{
Five field low-surface brightness galaxies in this analysis.
Here,
 $M_{\rm dyn}$ and $R_c$ denote the dynamical mass and core radius, respectively,
 as determined from rotation curve in \cite{DiPaolo:2019eib}.
The low values of the ratio $M_{\rm HI}/M_{\rm dyn}$
 indicate that very low quantity of baryon included in there galaxies.
}
\label{five-LSBGs}
\end{center}
\end{table}

We first consider the case
 in which these five galaxies decoupled from the proto--cluster at $z=2.45$
 \cite{Diener:2014efa}.
The initial density parameter in the NFW profile is estimates as
\begin{equation}
 \rho_s = A \times (1+\delta) \, \rho_m(z),
\end{equation}
 where $A \simeq 100$, $\delta \simeq 10$, $z=2.45$ and
 $\rho_m(z)=\rho_{\rm crit,0} \, \Omega_m(1+z)^3$.
This value is taken to be the same for all five galaxies.
The corresponding values of $r_s$ are obtained
 by solving eq.(\ref{mass-explicit})
 using $M_{\rm mem}=M_{\rm dyn}$ and $R_{\rm mem}=R_c$
 for each five galaxy.
Using these parameters,
 we compute the core formation timescale $t_{\rm core}$
 from eq.(\ref{time-core-form}) and eq.(\ref{initial-tr}).
There timescales should be compared with
 the cosmic time of approximately $11$ Gyr corresponding to $z=2.45$.

The results for
 $\sigma = 1 \, {\rm cm}^2/{\rm g}$ and $\sigma = 0.1 \, {\rm cm}^2/{\rm g}$
 are shown tables \ref{high-sigma-2.45} and \ref{low-sigma-2.45}, respectively.
For $\sigma = 1 \, {\rm cm}^2/{\rm g}$
 the core times are much shorter than $11$ Gyr,
 and in most cases the system would undergo gravothermal collapse in the black holes
 before the present time (except for F563-V1 and marginally F574-2).
We therefore conclude that $\sigma = 1 \, {\rm cm}^2/{\rm g}$ is too large.
In contrast, for $\sigma = 0.1 \, {\rm cm}^2/{\rm g}$
 the values of $t_{\rm core}$ are comparable to, or slightly smaller than $11$ Gyr.
Given that $t_{\rm core} \equiv 30 \times t_{r,0}$
 is only an approximate definition as a reference
 and that the profile remain stable up to $8 \times t_{\rm core}$
 (see section \ref{sec:SIDM}),
 these results are consistent with the presence of cores at the present time.
 
\begin{table}[t]
\begin{center}
\large
\begin{tabular}{|l|c|c|c|c|} \hline
 name    & $A$ & $\rho_s$ [$M_\odot/{\rm kpc}^3]$ & $r_s$ [kpc] & $t_{\rm core}$ [Gyr]
 \\ \hline\hline
 F561-1  & $100$ & $1.8\times10^6$ & 6.2 & 0.54 \\ \hline
 F563-V1 & $100$ & $1.8\times10^6$ & 3.9 & 0.87 \\ \hline
 F571-V1 & $100$ & $1.8\times10^6$ & 11  & 0.30 \\ \hline
 F574-1  & $100$ & $1.8\times10^6$ & 13  & 0.27 \\ \hline
 F574-2  & $100$ & $1.8\times10^6$ & 4.9 & 0.69 \\ \hline
\end{tabular}
\caption{
Results for $\sigma = 1 \, {\rm cm}^2/{\rm g}$ for a proto--cluster at $z=2.45$.
All values of $t_{\rm core}$ are too small,
 indicating that black holes would form before the present time.
}
\label{high-sigma-2.45}
\end{center}
\end{table}
\begin{table}[t]
\begin{center}
\large
\begin{tabular}{|l|c|c|c|c|} \hline
 name    & $A$ & $\rho_s$ [$M_\odot/{\rm kpc}^3]$ & $r_s$ [kpc] & $t_{\rm core}$ [Gyr]
 \\ \hline\hline
 F561-1  & $100$ & $1.8\times10^6$ & 6.2 & 5.4 \\ \hline
 F563-V1 & $100$ & $1.8\times10^6$ & 3.9 & 8.7 \\ \hline
 F571-V1 & $100$ & $1.8\times10^6$ & 11  & 3.0 \\ \hline
 F574-1  & $100$ & $1.8\times10^6$ & 13  & 2.7 \\ \hline
 F574-2  & $100$ & $1.8\times10^6$ & 4.9 & 6.9 \\ \hline
\end{tabular}
\caption{
Results for $\sigma = 0.1 \, {\rm cm}^2/{\rm g}$ for a proto--cluster at $z=2.45$.
}
\label{low-sigma-2.45}
\end{center}
\end{table}

We next consider the case where
 the galaxies decoupled at proto--cluster at $z=7.66$ \cite{Laporte:2022},
 corresponding to a cosmic time of approximately $13$ Gyr.
The results are shown in tables \ref{high-sigma-7.66} and \ref{low-sigma-7.66}.
As in the previous case,
 for $\sigma = 1 \, {\rm cm}^2/{\rm g}$,
 most values of $t_{\rm core}$ are too small,
 leading to black hole formation before the present time (with minor exceptions).
This again indicates that such a large cross section is disfavored.
For $\sigma = 0.1 \, {\rm cm}^2/{\rm g}$,
 the resulting values of $t_{\rm core}$
 are the same order as, or slightly smaller than, $13$ Gyr.
Since the core structure remains stable for times up to $\sim 8,t_{\rm core}$,
 these values are consistent with core formation at the present epoch.

\begin{table}[t]
\begin{center}
\large
\begin{tabular}{|l|c|c|c|c|} \hline
 name    & $A$ & $\rho_s$ [$M_\odot/{\rm kpc}^3]$ & $r_s$ [kpc] & $t_{\rm core}$ [Gyr]
 \\ \hline\hline
 F561-1  & $10$ & $1.3\times10^6$ & 7.3 & 0.76 \\ \hline
 F563-V1 & $10$ & $1.3\times10^6$ & 4.5 & 1.2  \\ \hline
 F571-V1 & $10$ & $1.3\times10^6$ & 13  & 0.41 \\ \hline
 F574-1  & $10$ & $1.3\times10^6$ & 15  & 0.37 \\ \hline
 F574-2  & $10$ & $1.3\times10^6$ & 5.6 & 0.99 \\ \hline
\end{tabular}
\caption{
Results for $\sigma = 1 \, {\rm cm}^2/{\rm g}$ for a proto--cluster at $z=7.66$.
All values of $t_{\rm core}$ are too small,
 indicating that black holes would form before the present time.
}
\label{high-sigma-7.66}
\end{center}
\end{table}
\begin{table}[t]
\begin{center}
\large
\begin{tabular}{|l|c|c|c|c|} \hline
 name    & $A$ & $\rho_s$ [$M_\odot/{\rm kpc}^3]$ & $r_s$ [kpc] & $t_{\rm core}$ [Gyr]
 \\ \hline\hline
 F561-1  & $10$ & $1.3\times10^6$ & 7.3 & 7.6 \\ \hline
 F563-V1 & $10$ & $1.3\times10^6$ & 4.5 & 12  \\ \hline
 F571-V1 & $10$ & $1.3\times10^6$ & 13  & 4.1 \\ \hline
 F574-1  & $10$ & $1.3\times10^6$ & 15  & 3.7 \\ \hline
 F574-2  & $10$ & $1.3\times10^6$ & 5.6 & 9.9 \\ \hline
\end{tabular}
\caption{
Results for $\sigma = 0.1 \, {\rm cm}^2/{\rm g}$ for a proto--cluster at $z=7.66$.
}
\label{low-sigma-7.66}
\end{center}
\end{table}

Finally,
 we consider the case of $z=7.88$ \cite{Morishita:2022}
 corresponding to the time about $13$ Gyr.
The results are given in tables \ref{high-sigma-7.88} and \ref{low-sigma-7.88}.
For $\sigma = 1 \, {\rm cm}^2/{\rm g}$,
 core formation occurs too rapidly,
 and most systems would collapse into black holes before the present time.
This again suggests that such a large cross section is not viable. 
For $\sigma = 0.1 \, {\rm cm}^2/{\rm g}$,
 the values of $t_{\rm core}$ are again comparable to, or somewhat smaller than,
 $13$ Gyr.
Considering the slow evolution of the core profile,
 these results support the formation of cores in the present universe.

\begin{table}[t]
\begin{center}
\large
\begin{tabular}{|l|c|c|c|c|} \hline
 name    & $A$ & $\rho_s$ [$M_\odot/{\rm kpc}^3]$ & $r_s$ [kpc] & $t_{\rm core}$ [Gyr]
 \\ \hline\hline
 F561-1  & $5$ & $3.6\times10^6$ & 4.7 & 0.27 \\ \hline
 F563-V1 & $5$ & $3.6\times10^6$ & 2.9 & 0.44 \\ \hline
 F571-V1 & $5$ & $3.6\times10^6$ & 8.2 & 0.15 \\ \hline
 F574-1  & $5$ & $3.6\times10^6$ & 9.3 & 0.14 \\ \hline
 F574-2  & $5$ & $3.6\times10^6$ & 3.7 & 0.34 \\ \hline
\end{tabular}
\caption{
Results for $\sigma = 1 \, {\rm cm}^2/{\rm g}$ for a proto--cluster at $z=7.88$.
All values of $t_{\rm core}$ are too small,
 indicating that black holes would form before the present time.
}
\label{high-sigma-7.88}
\end{center}
\end{table}
\begin{table}[t]
\begin{center}
\large
\begin{tabular}{|l|c|c|c|c|} \hline
 name    & $A$ & $\rho_s$ [$M_\odot/{\rm kpc}^3]$ & $r_s$ [kpc] & $t_{\rm core}$ [Gyr]
 \\ \hline\hline
 F561-1  & $5$ & $3.6\times10^6$ & 4.7 & 2.7 \\ \hline
 F563-V1 & $5$ & $3.6\times10^6$ & 2.9 & 4.4 \\ \hline
 F571-V1 & $5$ & $3.6\times10^6$ & 8.2 & 1.5 \\ \hline
 F574-1  & $5$ & $3.6\times10^6$ & 9.3 & 1.4 \\ \hline
 F574-2  & $5$ & $3.6\times10^6$ & 3.7 & 3.4 \\ \hline
\end{tabular}
\caption{
Results for $\sigma = 0.1 \, {\rm cm}^2/{\rm g}$ for a proto--cluster at $z=7.88$.
}
\label{low-sigma-7.88}
\end{center}
\end{table}

In summary,
 we find that $\sigma = 1 \, {\rm cm}^2/{\rm g}$ is too large,
 while $\sigma = 0.1 \, {\rm cm}^2/{\rm g}$ provides a consistent explanation
 for core formation across all three proto--cluster scenarios.
Although our analysis
 is based on order-of-magnitude estimates and several assumptions,
 the consistency of the results across different proto--clusters is notable.
It is also interesting
 that the estimated values of $\rho_s$
 are similar across the three proto--clusters,
 while the values of $r_s$ show some variation.
Since $\sigma = 0.1 \, {\rm cm}^2/{\rm g}$
 is consistent with constraints from the shapes of dark matter halos
 in present--day clusters,
 velocity--dependent self--interaction may not be required.

\section{Conclusions}
\label{sec:conclusions}

We have investigated the possibility
 that gravothermal collapse driven by self--interacting dark matter
 contributes the formation of the dark matter core in galaxies.
Since the violent baryonic effects and major galactic mergers
 can obscure the impact of gravothermal collapse as a slow thermal process,
 we focus on galaxies in relatively ``silent'' environment,
 namely low--surface brightness galaxies (LSBGs) in the field.
We assume that
 such LSBGs were once member of proto--cluster and later migrated into the field.
This is a strong assumption and
 should be tested by future simulations of structure formation.
Under this assumption
 we can estimate the initial conditions for galaxy evolution
 using the observational information on proto--cluster at the corresponding redshift.
Because the LSBGs contain very small amounts of baryonic matter,
 we estimate the timescale required for core formation
 driven solely self--interacting dark matter 
 based on the analyses in \cite{Pollack:2014rja,Choquette:2018lvq}.
We consider three observed proto--clusters
 at $z=2.45$, $z=7.66$ and $z=7.88$,
 and compare the required core formation timescales
 for five specific field LSBGs studied in \cite{DiPaolo:2019eib,deBlok:1996jib}.
A common result across all three proto--clusters is as follows.
A self--interaction cross section of $\sigma = 1 \, {\rm cm}^2/{\rm g}$
 is too large and leads to black hole formation before the present time.
No such black holes have been observed in our sample of five galaxies.
In contrast
 a cross section of $\sigma = 0.1 \, {\rm cm}^2/{\rm g}$
 is consistent with the formation of cored density profiles at the present time.
Although our analysis is based on order-of-magnitude estimates
 and relies on several assumptions,
 the consistency of the results across all three proto--clusters is noteworthy.
This suggests that velocity--dependent self--interaction may not be necessary,
 since the value $\sigma = 0.1 \, {\rm cm}^2/{\rm g}$
 is also consistent with constraints derived
 from shapes of present--day cluster helos.

If the dark matter is described by a real scalar field $\phi(x)$
 the presence of self--interactions is natural
 from the viewpoint of renormalizability.
It is also natural to include
 a coupling to the standard model Higgs doublet $\Phi(x)$
 of the form $\Phi^{\dag}\Phi \phi^2$,
 corresponding to a Higgs portal model with an ``integrating--in'' mechanism
 for dark matter production \cite{Hall:2009bx}.
Assuming a perturbative quartic self--interaction $\phi^4$,
 the dark matter mass is expected to be of the order of $10$ MeV.
The more detailed study of the production of self--interacting dark matter
 is left for future work.

\section*{Acknowledgments}

This study was supported
 by the Basic Research Fund of Tokyo Metropolitan University.


\begin{thebibliography}{99}


\bibitem{Flores:1994gz}
R.~A.~Flores and J.~R.~Primack,
``Observational and theoretical constraints on singular dark matter halos,''
Astrophys. J. Lett. \textbf{427} (1994), L1-4
[arXiv:astro-ph/9402004 [astro-ph]].
\bibitem{Moore:1994yx}
B.~Moore,
``Evidence against dissipationless dark matter from observations of galaxy haloes,''
Nature \textbf{370} (1994), 629.

\bibitem{Navarro:1996gj}
J.~F.~Navarro, C.~S.~Frenk and S.~D.~M.~White,
``A Universal density profile from hierarchical clustering,''
Astrophys. J. \textbf{490} (1997), 493-508
[arXiv:astro-ph/9611107 [astro-ph]].

\bibitem{Lynden-Bell:1968eqn}
D.~Lynden-Bell and R.~Wood,
``The gravo-thermal catastrophe in isothermal spheres
 and the onset of red-giant structure for stellar systems,''
Mon. Not. Roy. Astron. Soc. \textbf{138} (1968), 495.
\bibitem{Lynden-Bell:1980xip}
D.~Lynden-Bell and P.~P.~Eggleton,
``On the consequences of the gravothermal catastrophe,''
Mon. Not. Roy. Astron. Soc. \textbf{191} (1980) no.3, 483-498.

\bibitem{Pollack:2014rja}
J.~Pollack, D.~N.~Spergel and P.~J.~Steinhardt,
``Supermassive Black Holes from Ultra-Strongly Self-Interacting Dark Matter,''
Astrophys. J. \textbf{804} (2015) no.2, 131
[arXiv:1501.00017 [astro-ph.CO]].
\bibitem{Choquette:2018lvq}
J.~Choquette, J.~M.~Cline and J.~M.~Cornell,
``Early formation of supermassive black holes via dark matter self-interactions,''
JCAP \textbf{07} (2019), 036
[arXiv:1812.05088 [astro-ph.CO]].

\bibitem{Koda:2011yb}
J.~Koda and P.~R.~Shapiro,
``Gravothermal collapse of isolated self-interacting dark matter haloes:
 N-body simulation versus the fluid model,''
Mon. Not. Roy. Astron. Soc. \textbf{415} (2011), 1125
[arXiv:1101.3097 [astro-ph.CO]].
\bibitem{Rocha:2012jg}
M.~Rocha, A.~H.~G.~Peter, J.~S.~Bullock, M.~Kaplinghat, S.~Garrison-Kimmel,
 J.~Onorbe and L.~A.~Moustakas,
``Cosmological Simulations with Self-Interacting Dark Matter I:
 Constant Density Cores and Substructure,''
Mon. Not. Roy. Astron. Soc. \textbf{430} (2013), 81-104
[arXiv:1208.3025 [astro-ph.CO]].
\bibitem{Elbert:2014bma}
O.~D.~Elbert, J.~S.~Bullock, S.~Garrison-Kimmel, M.~Rocha, J.~O{\~n}orbe
 and A.~H.~G.~Peter,
``Core formation in dwarf haloes with self-interacting dark matter:
 no fine-tuning necessary,''
Mon. Not. Roy. Astron. Soc. \textbf{453} (2015) no.1, 29-37
[arXiv:1412.1477 [astro-ph.GA]].

\bibitem{Tulin:2017ara}
S.~Tulin and H.~B.~Yu,
``Dark Matter Self-interactions and Small Scale Structure,''
Phys. Rept. \textbf{730} (2018), 1-57
[arXiv:1705.02358 [hep-ph]].

\bibitem{DiPaolo:2019eib}
C.~Di Paolo, P.~Salucci and A.~Erkurt,
``The universal rotation curve of low surface brightness galaxies {\textendash} IV.
 The interrelation between dark and luminous matter,''
Mon. Not. Roy. Astron. Soc. \textbf{490} (2019) no.4, 5451-5477.

\bibitem{Diener:2014efa}
C.~Diener, S.~Lilly, C.~Ledoux, G.~Zamorani, D.~Murphy, M.~Bolzonella,
 P.~Capak, O.~Ilbert and H.~McCracken,
``A proto-cluster at z=2.45,''
Astrophys. J. \textbf{802} (2015), 31
[arXiv:1411.0649 [astro-ph.CO]].
\bibitem{Laporte:2022}
N.~Laporte, A.~Zitrin, H.~Dole, G.=Roberts-Borsani, L.J.~Furtak, C.~Witten,
``A lensed protocluster candidate at z=7.66 identified in JWST observations
 of the galaxy cluster SMACS0723-7327,''
Astron. Astrophys. \textbf{667} (2022) L3
[arXiv:2208.04930 [astro-ph.GA]].
\bibitem{Morishita:2022}
T.~Morishita at al.,
``Early results from GLASS-JWST. XIV: A spectroscopically confirmed protocluster
 650 million years after the Big Bang,''
Astrophys. J. \textbf{947} (2023) L24
[arXiv:2211.09097 [astro-ph.GA]].

\bibitem{ParticleDataGroup:2024cfk}
S.~Navas \textit{et al.} [Particle Data Group],
``Review of particle physics,''
Phys. Rev. D \textbf{110} (2024) no.3, 030001.

\bibitem{Springel:2005nw}
V.~Springel, S.~D.~M.~White, A.~Jenkins, C.~S.~Frenk, N.~Yoshida, L.~Gao,
 J.~Navarro, R.~Thacker, D.~Croton and J.~Helly, \textit{et al.}
``Simulating the joint evolution of quasars, galaxies
 and their large-scale distribution,''
Nature \textbf{435} (2005), 629-636
[arXiv:astro-ph/0504097 [astro-ph]].
\bibitem{Kitzbichler:2006ec}
M.~G.~Kitzbichler and S.~D.~M.~White,
``The high redshift galaxy population in hierarchical galaxy formation models,''
Mon. Not. Roy. Astron. Soc. \textbf{376} (2007), 2-12
[arXiv:astro-ph/0609636 [astro-ph]].
\bibitem{Henriques:2011xn}
B.~Henriques, S.~White, G.~Lemson, P.~Thomas, Q.~Guo, G.~D.~Marleau and R.~Overzier,
``Confronting theoretical models
 with the observed evolution of the galaxy population out to z=4,''
Mon. Not. Roy. Astron. Soc. \textbf{421} (2012), 2904
[arXiv:1109.3457 [astro-ph.CO]].


\bibitem{Balberg:2002ue}
S.~Balberg, S.~L.~Shapiro and S.~Inagaki,
``Selfinteracting dark matter halos and the gravothermal catastrophe,''
Astrophys. J. \textbf{568} (2002), 475-487
[arXiv:astro-ph/0110561 [astro-ph]].


\bibitem{deBlok:1996jib}
W.~J.~G.~de Blok, S.~S.~McGaugh and J.~M.~van der Hulst,
``Hi observations of low surface brightness galaxies: probing low density galaxies,''
Mon. Not. Roy. Astron. Soc. \textbf{283} (1996), 18-54
[arXiv:astro-ph/9605069 [astro-ph]].


\bibitem{Hall:2009bx}
L.~J.~Hall, K.~Jedamzik, J.~March-Russell and S.~M.~West,
``Freeze-In Production of FIMP Dark Matter,''
JHEP \textbf{03} (2010), 080
[arXiv:0911.1120 [hep-ph]].
\end{thebibliography}
\end{document}